\documentclass[%
 reprint,
 amsmath,amssymb,
 aps,
]{revtex4-2}

\usepackage{graphicx}
\usepackage{dcolumn}
\usepackage{bm}
\newcommand{\beq}{\begin{eqnarray}}
\newcommand{\eeq}{\end{eqnarray}}
\usepackage{comment}

\begin{document}

\preprint{}

\title{Single-point statistical moments of the nonhomogeneous stochastic advection equation in the small correlation length limit}
\thanks{The corresponding author is grateful to R. Stephenson for a technical suggestion that was instrumental to the derivation in Sec. III.}%

\author{Keiko Kircher}
\author{Cristian Proistosescu}
\author{Ryan L. Sriver}
 \altaffiliation[]{Department of Atmospheric Sciences, University of Illinois at Urbana-Champaign, 1301 W Green St, Urbana, IL 61801, USA}
 \email{keikoino@illinois.edu}




\date{\today}

\begin{abstract}
This paper presents the derivation of closed-form expressions of the single-point statistical moments of a solution to a nonhomogeneous stochastic advection equation with a linear relaxation. While analytical solutions exist for homogeneous systems, nonhomogeneous cases have traditionally relied on intensive numerical simulations. Here, we provide an analytical framework for calculating single-point statistical moments by first obtaining the solution to the stochastic advection equation via the method of characteristics, from which the moments are derived.  Explicit, closed-form expressions for the first four moments are derived as functions of the characteristic length scale of the stochastic velocity field and the spatial derivatives of time-mean profile of the field. The analytical results are validated against numerical simulations, demonstrating excellent agreement across a range of physical parameters. The resulting theory acts as a generalized ``equation-of-state" style approach for predicting variability and non-Gaussian statistical behavior directly from the macroscopic mean state, providing applicability across transport systems with a wide range of time and length scales, including geology, hydrology, and atmospheric sciences.
\end{abstract}

\maketitle


\section{\label{sec: introduction}Introduction}

The advection equation is a fundamental tool for characterizing the transport of physical quantities within fluid media. It is extensively utilized to model tracer transport such as advection of chemical concentration in porous media \cite{selvadurai2004advective, hunt2017flow, yu2019hybrid}, the evolution of scalar fields such as temperature \cite{jochum2006temperature}, humidity \cite{raymond2000moisture}, and pollutant concentrations \cite{scott1997particle, zhang2022backward} in the high-P\'{e}clet-number regimes where diffusion effects become negligible.

While certain physical systems are adequately described by deterministic velocity fields, others are more accurately represented by stochastic transport velocities. In geology and hydrology, the advection equation with a random velocity field describes solute transport with random permeabilities in porous media \cite{dagan1984solute, mauri2003heat, kondratenko2007random, xu2017method}. In atmospheric sciences, it describes transport of passive scalars such as temperature in highly turbulent regions \cite{linz2018large, tamarin2022simple,kircher2025analytical}. On astrophysical scales, similar stochastic ideas can be applied to model the propagation of cosmic rays, where the turbulent magnetic fields induce random particle trajectories \cite{giacalone1999transport, supanitsky2021cosmic}. 

Consequently, significant efforts have been directed toward understanding the advection of physical quantities with random velocity fields to elucidate the statistical properties of the resulting solutions. Following Kolmogorov's work \cite{kolmogorov1991dissipation}, Kraichnan \cite{kraichnan1994anomalous} applied the idea of structure functions to passive scalars advected by random velocity fields to calculate anomalous scaling of structure functions. Substantial work has  been done to refine Kraichnan's theory, such as calculating more precise scaling exponents \cite{yakhot1997passive}, consideration of a finite correlation time \cite{eyink1996intermittency}, and combining the theory with Taylor's frozen-flow model \cite{wang2025solvable}. Dynamical considerations have been investigated through equations of motion with an additive random forcing term to calculate the system's PDF via the Langevin and Fokker-Planck frameworks \cite{aringazin2004one}, as well as through direct approaches to solving the advection equation with random velocity fields \cite{dorini2007finite, dorini2009evaluation, santos2010probability, calatayud2020extending}. 


Although these previous studies have significantly advanced the stochastic transport theory, their analytical frameworks focused primarily on homogeneous advection equations or on multi-point observables, most notably structure functions. While structure functions are effective for determining scale-invariant properties by calculating the moments of the spatial difference of a field, they do not contain any information about statistical moments at a fixed location. However, tracking these local single-point statistics is critical for practical risk assessment, as they determine the frequency and predictability of extreme localized phenomena, such as extreme urban temperature events or acute pollution episodes. As a result, calculation of single-point statistical moments of the solution to a nonhomogeneous advection equation (i.e., with nonzero source terms) remained heavily dependent on numerical simulations \cite{ginzburg2007lattice, xu2017method, alqahtani2022extreme, linz2020framework, tamarin2020changes}. 

In this study, we present a fully analytical framework for calculating the single-point statistical moments of the solution of the nonhomogeneous advection equation subject to stochastic transport and linear relaxation, where the linear relaxation mechanism models a continuous source and sink that drives the field back toward its spatial equilibrium profile. To derive closed-form expressions, we adopt two idealized assumptions: first, the velocity field is modeled as a purely time-dependent stochastic function, changing its values at every fixed interval as a sequence of random step functions; second, the average displacement of the field within a single correlation time is negligible relative to the spatial length scale of the change in equilibrium profile. Under these constraints, we derive closed-form solutions that are dependent solely on the statistical properties of the velocity fields and the spatial derivatives of the mean profile of the field. By connecting moments with the mean profile with closed-form expressions, our framework provides a highly efficient alternative to traditional numerical simulations. More fundamentally, these analytical expressions establish an ``equation-of-state" style approach for the transport system, directly mapping microscopic temporal variability (the localized statistical moments) onto macroscopic spatial structure (the mean profile and its derivatives).


\section{\label{sec: PDE}Model Setup}

\subsection{Stochastic Advection Equation and Its Exact Solution}

Let us consider a one-dimensional fluid that transports a field $u$, where the transport of $u$ occurs stochastically due to a random velocity field $v(t)$ while approaching an equilibrium profile $u_e$ over time. As an idealized approximation of this process, we use a one-dimensional stochastic advection equation with linear relaxation:
\beq
\frac{\partial u(x,t)}{\partial t} + v(t)\frac{\partial u(x,t)}{\partial x} = -\frac{u(x,t)-u_e(x)}{\tau}, \label{eq: advection}
\eeq
where $\tau$ is the characteristic timescale of the linear relaxation, $u_e(x)$ is assumed to be a smooth function that depends only on one variable $x$, and $v(t)$ is a Gaussian noise that depends only one variable $t$.  

Note that Eq. (\ref{eq: advection}) is a semilinear PDE, hence it can be solved via the method of characteristics, as was done by \cite{dorini2011linear} in the homogeneous case. The characteristic equations of Eq. (\ref{eq: advection}) are
\beq
\frac{dt}{dr} = 1 \\
\frac{dx}{dr} = v(r) \\
\frac{du}{dr} = -\frac{u-u_e(x)}{\tau}.
\eeq


With the initial conditions $t(0) = 0$, $x(0) = s$, and $u(0,s) = u_i(s)$, we obtain 
\beq
&r=t& \label{eq: t}\\
&x = s + \Delta s (r)& \label{eq: phi}\\
&\Delta s (r) = \int _0 ^r v(r') dr'& \label{eq: Deltas},
\eeq

\beq
\begin{split}
&u(r,s) = u_i(s)e^{-r/\tau} &\\
&+e^{-r/\tau}\frac 1\tau \int_0 ^r u_e(s+\Delta s(r'))e^{r'/\tau} dr'.& \label{eq: solution_general}
\end{split}
\eeq
Since $u_e(s)$ is differentiable with respect to $s$ and $\Delta s(r')$ is differentiable in $r'$, we now rewrite Eq. (\ref{eq: solution_general}) with integration by parts:
\beq
\begin{split}
&u(r,s) = u_i(s) e^{-r/\tau} + u_e(s+\Delta s) - u_e(s)e^{-r/\tau}&\\
& - e^{-r/\tau}\int_0 ^r \frac{\partial u_e}{\partial s}(s+\Delta s(r')) v(r')e^{r'/\tau} dr'. &\label{eq: soln_after_integ_by_parts_1}
\end{split}
\eeq
In the case of $\tau \ll t$, Eq. (\ref{eq: soln_after_integ_by_parts_1}) approaches 
\beq
\begin{split}
&u(r,s) = u_e(s+\Delta s)&\\
& - e^{-r/\tau}\int_0 ^r \frac{\partial u_e}{\partial s}(s+\Delta s(r')) v(r')e^{r'/\tau} dr', &\label{eq: soln_after_integ_by_parts}
\end{split}
\eeq
which is clearly independent of the initial condition (and coincides with the exact solution when $u(0,s) = u_e(s)$ is used as the initial condition). 

Equation (\ref{eq: soln_after_integ_by_parts}), along with Eq. (\ref{eq: t})--(\ref{eq: phi}), provides the exact steady-state solution to Eq. (\ref{eq: advection}) for any velocity field $v(t)$ regardless of the initial condition. However, calculating the statistical moments of $u(t,x)$ in this general form remains analytically intractable due to the complexity of the integral.

\subsection{\label{subsec: limits}Asymptotic Expansion and Noise Discretization}

To obtain a closed-form solution of statistical moments of $u(t,x)$, we discretize the noise $v(t)$ with a series of step functions that take a random value at each step with a Gaussian distribution (conceptually mimicking a staircase with random step heights in time), given by
\beq
v(t) =  \sum _i v_i \left( H(t - t_i) - H(t - (t_i+\Delta t)) \right), \label{eq: v}
\eeq
where $v_i \sim \mathcal{N}(0,\sigma _v^2)$, $v_i$ and $v_j$ are uncorrelated if $i\ne j$, and $H(x)$ is a Heaviside step function, where we use the convention $H(0) = 1/2$.  From now on, we shall call each step of such 'stairs' a time step for convenience. A collection of step functions represents simplified noise with a nonzero autocorrelation, where the usual exponential decay of the autocorrelation is replaced by a constant value. In this study, we consider only the case where $\Delta t \ll \tau$: when the timescale of the relaxation to the average profile $u_e$ is much larger than the timescale of the autocorrelation of the noise $\Delta t$. In addition, we restrict our analysis to the regime where 
\beq
v(t) \Delta t \ll \frac{d ^n u_e(x)/dx ^n}{d^{n+1} u_e(x)/d
x^{n+1}}, \ n = 0,1, 2, \dots  \label{eq:regime}
\eeq
on average; this is the regime where the length scale of the spatial change in the $n^{\text{th}}$ derivative of the equilibrium profile is much larger than the typical distance the fluid moves within the timescale of the noise $\Delta t$. Mathematically, these approximations let terms of the lowest order in $v(t) \Delta t$ (as well as $\Delta x =  \int _0 ^t v(t'') dt''$) and of the lowest order in the derivative of $u_e$ dominate. In fact, since the order of the derivative of $u_e$ increases as the order of  $v(t) \Delta t$ increases, it is sufficient to keep the terms that are the lowest order in $v$. This is a regime applicable to physical systems such as Earth's atmospheres \cite{taylor1922diffusion, bennett1987lagrangian, Daoud:2003, Swanson:1997, schneider2015physics}, and its validity is further demonstrated in Sec. \ref{sec: validation}. 

Utilizing this limit, we expand $\partial u_e/\partial s(s+\Delta s)$ in the integrand in Eq.(\ref{eq: soln_after_integ_by_parts}) about $s$:
\beq
\begin{split}
&u(r,s) = u_e(s+\Delta s) -  u_e'(s)e^{-r/\tau} \int_0 ^r  v(r')e^{r'/\tau} dr' &\\
&- u_e''(s)e^{-r/\tau} \int_0 ^r \Delta s(r') v(r') e^{r'/\tau} dr' + \dots, & \label{eq: soln3_1}
\end{split}
\eeq
where a prime is used to indicate the derivative with respect to its argument (i.e., $u_e'(s) =  du_e(s)/ds$). Now Eq. (\ref{eq: soln3_1}) can be expressed in terms of $t$ and $x$ via Eqs. (\ref{eq: t})--(\ref{eq: Deltas}) as
\beq
\begin{split}
&u(t,x) = u_e(x)- u_e'(x - \Delta x)  e^{-t/\tau} \int_0 ^t  v(t')e^{t'/\tau} dt' &\\
&- u_e''(x-\Delta x)  e^{-t/\tau} \int_0 ^t \Delta s(t') v(t') e^{t'/\tau} dt' - \dots & \label{eq: soln3}
\end{split}
\eeq
We further expand this expression about $x$ to find
\beq
\begin{split}
&u(t, x) = u_e(x) - u_e'(x)e^{-t/\tau} \int_0 ^t  v(t')e^{t'/\tau} dt'  &\\
&-u_e''(x)e^{-t/\tau}\int_0 ^t (\Delta x(t') v(t') - \Delta x (t) v(t'))e^{t'/\tau} dt' &\\
&- \dots . &\label{eq: soln4}
\end{split}
\eeq
Equation (\ref{eq: soln4}) and our asymptotic limits serve as the basis for calculating analytical forms of statistical moments of the solution $u(t, x)$ in the next section. 

\section{\label{sec: derivation}Statistical Moments}

We now proceed with the step-by-step derivation of the first four statistical moments in terms of statistical properties of the velocity fields and the spatial derivatives of the mean profile $\bar u$, using the asymptotic limit established in Subsec. \ref{subsec: limits}. For conciseness, the explicit spatial argument $x$ in $u_e(x)$ and its derived quantities will be omitted hereafter.

\subsection{First Moment}

\begin{figure}
\centering
    \includegraphics[height = 4 cm]{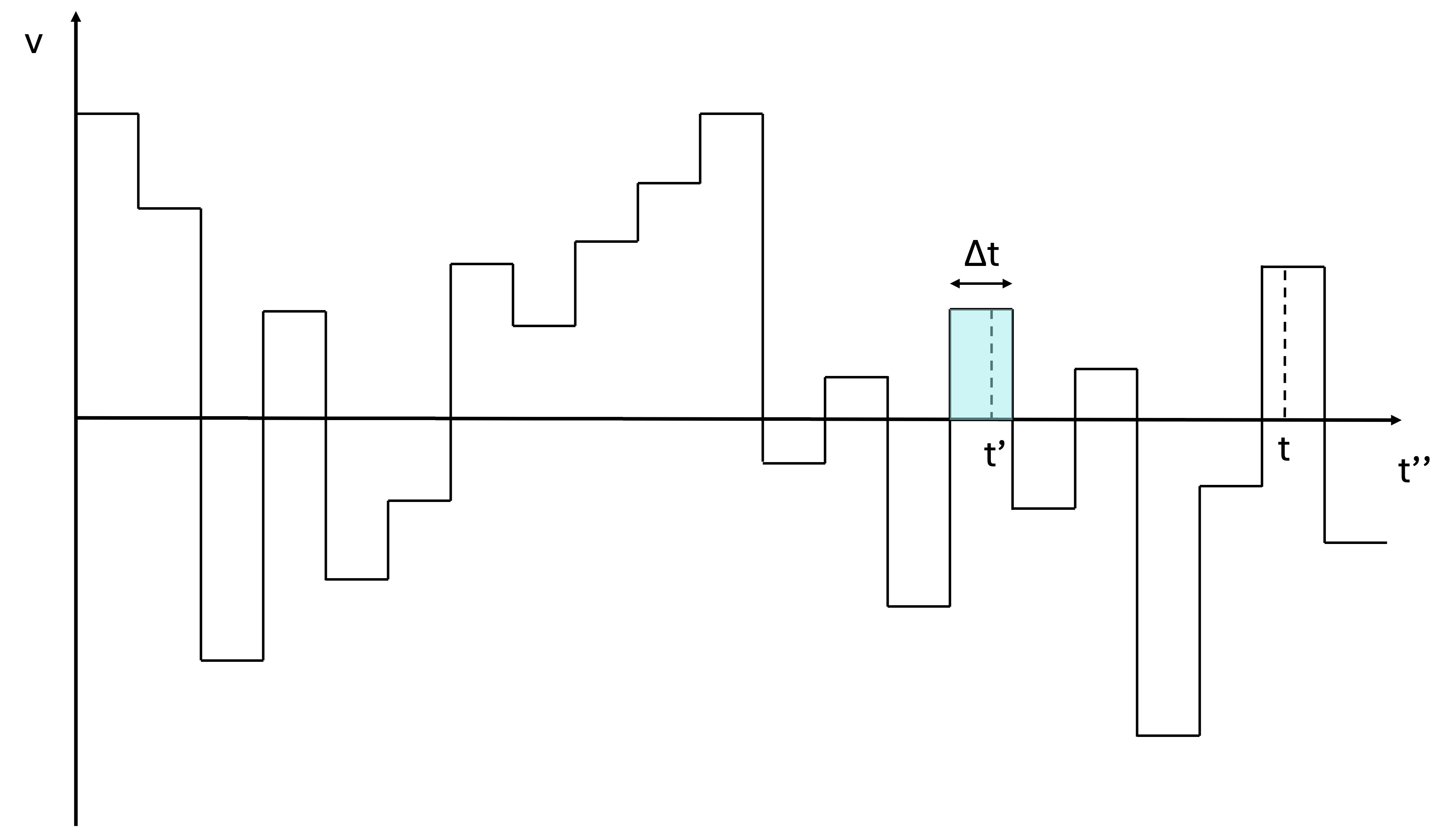}
    \caption{A schematic of the stochastic velocity given by Eq. (\ref{eq: v}) as a function of time. A nonzero correlation between $v(t')$ and $\Delta x (t)= \int _0 ^t v(t'') dt''$ for $t' \le t$ occurs only at the time step that includes $t'' = t'$, and the correlation is present for the entire time step $\Delta t$.}
    \label{figure: phi_t_v_correlation}
\end{figure}

The first moment (average), evaluated to the lowest order in $v$, is given by taking the ensemble average of Eq. (\ref{eq: soln4}):
\beq
\begin{split}
&\bar u= u_e-u_e'' \int_0 ^t (\overline{\Delta x(t') v(t')} - \overline{\Delta x (t) v(t')})e^{(t'-t)/\tau} dt', &\label{eq: first_moment}
\end{split}
\eeq
where the bar indicates the ensemble average. We first consider the second term in the integrand, $\overline{\Delta x (t) v(t')}$. Since $t'\le t$, somewhere within the domain of integration for $v(t'')$ in $t''$ in the calculation of $\Delta x =  \int _0 ^t v(t'') dt''$, there exists a time step that includes $t'$. This is the only part of $\Delta x (t)$ that is correlated with $v(t')$, and unless $t'\le t$ is in the last time step of the integration, the entire time step is integrated in the calculation of $\Delta x (t)$ (see FIG. \ref{figure: phi_t_v_correlation}). We ignore the case where $t'$ is in the last time step, as despite its existence, the difference makes a negligible contribution to the integral in Eq.(\ref{eq: first_moment}) when $t \gg \tau \gg \Delta t$. Hence,
\beq
\overline{\Delta x(t) v(t')}  = \overline{v(t')^2} \Delta t = \sigma _v^2 \Delta t,
\eeq
where $\sigma _v$ is the standard deviation of $v(t)$.

\begin{figure}
\centering
    \includegraphics[height = 4 cm]{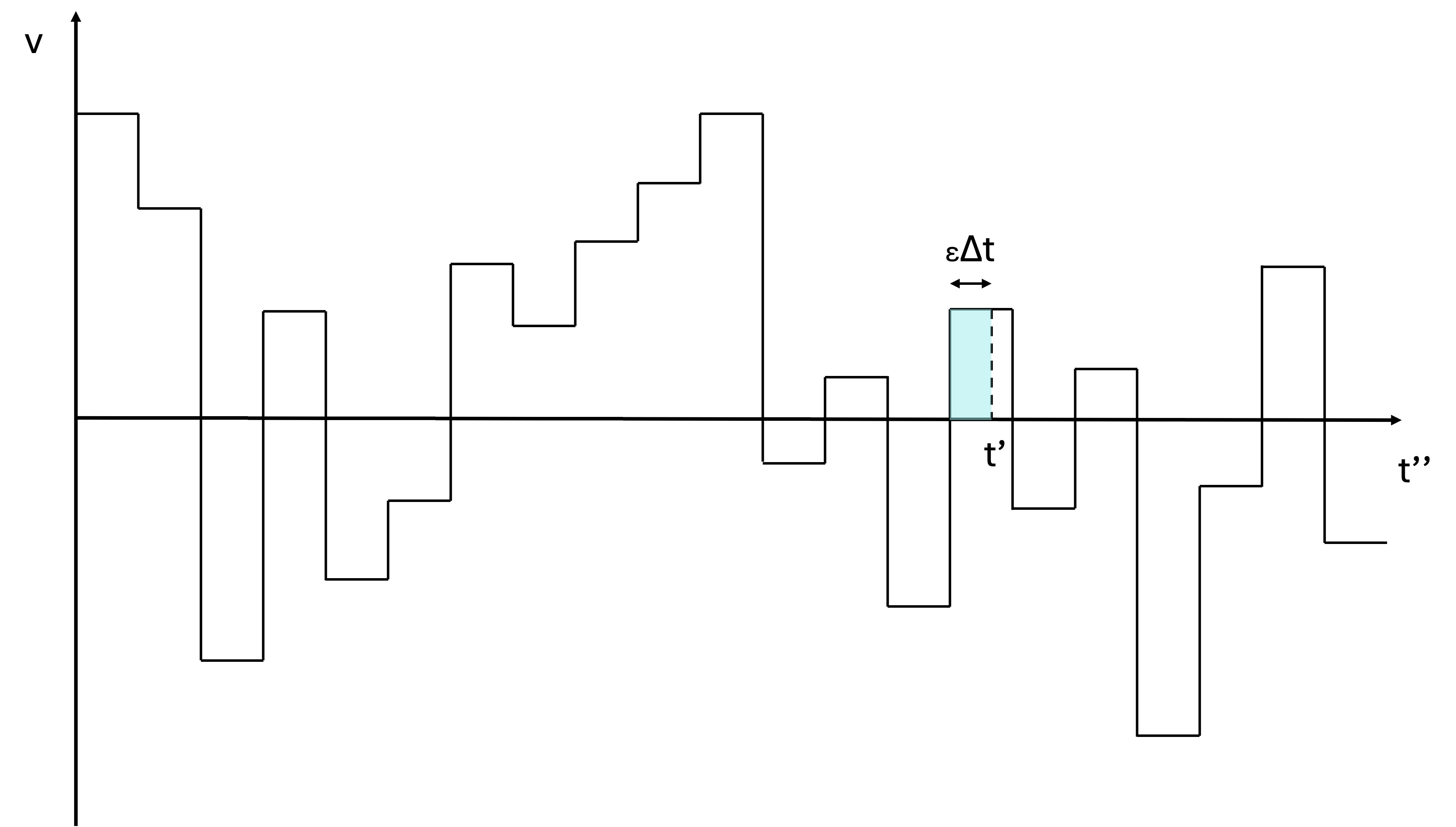}
    \caption{A schematic of the stochastic velocity given by Eq. (\ref{eq: v}) as a function of time. A nonzero correlation between $v(t')$ and $\Delta x (t')= \int _0 ^{t'} v(t'') dt''$ occurs only at the last time step of the integration, the one that includes $t'' = t'$, and the correlation is present only for a fraction $\epsilon$ of the time step $\Delta t$.}
    \label{figure: phi_tprime_v_correlation}
\end{figure}

On the other hand, the first term in the integrand of Eq. (\ref{eq: first_moment}) is given by 
\beq
\overline{\Delta x(t') v(t')}  = \epsilon (t') \sigma _v^2 \Delta t,
\eeq
where $\epsilon (t')$ is the fraction of the time step that includes $t'$. Since the integration of $v(t'')$ in $\Delta x (t')= \int _0 ^{t'} v(t'') dt''$ stops at $t'$, the nonzero correlation between $\Delta x (t')$ and $v(t')$ is present only within a fraction $\epsilon (t')$ of the time step that includes $t'$ (see FIG. \ref{figure: phi_tprime_v_correlation}).

Since $\epsilon (t')$ takes purely random values between 0 and 1 at each $t'$, its average value is 1/2. Thus, after the integration in $t'$ in Eq. (\ref{eq: first_moment}), we find the first moment
 \beq
 \bar u (x) = u_e + \frac 12 u_e'' \sigma _v ^2 \Delta t \tau = u_e + u_e'' \sigma_x ^2, \label{eq: moment1}
 \eeq
 where 
 \beq
 \sigma _x ^2 = \frac 12\sigma _v ^2 \Delta t \tau. \label{eq: sigma_phi}
 \eeq
Equation (\ref{eq: sigma_phi}) shows that the ensemble mean of $u$ (equivalent to time mean at a fixed location) is shifted from the equilibrium profile by a quantity proportional to the correlation time, relaxation time, variance of the velocity field, and square of the second derivative of the equilibrium profile. Thus, a linear $u_e$ does not shift $\bar u$ from $u_e$, but a nonlinear $u_e$ shifts the ensemble mean with the difference that increases linearly with both correlation time and relaxation time (in a similar way as a random walk, where relaxation time is taken as the total time passed). 

\subsection{Second Moment} 

 The second moment (variance), evaluated to the lowest order in $v$, is 
 \beq
 \begin{split}
 &\sigma _u ^2 = \overline{(u-\bar u)^2} &\\
 &= e^{-2t/\tau} u_e ^{\prime 2} \int _0 ^t \int_0 ^t \overline{v(t')v(t'')}e^{(t'+t'')/\tau} dt' dt'',&
 \end{split}
 \eeq
 where
 \beq
 \overline{v(t')v(t'')} = \sigma _v ^2  
 \eeq
 when $t''$ and $t'$ are in the same time step, and zero otherwise. Thus, in the limit $\tau \ll t$,
 \beq
 \begin{split}
 &\sigma _u^2 = e^{-2t/\tau}u_e^{\prime 2}  \int_0^t \sigma_v^2 \Delta t e^{2t'/\tau} dt' & \\
 &= \frac 12 u_e^{\prime 2}  \sigma _v ^2 \tau \Delta t = u_e^{\prime 2}  \sigma _x^2.&
 \end{split}
 \eeq
We now use Eq. (\ref{eq: moment1}) to rewrite the variance in terms of $\bar u(x)$ and keep only the lowest order in $\sigma_v$ to find
 \beq
 \sigma _u^2 = \bar u^{\prime 2}  \sigma _x^2. \label{eq: second_moment}
 \eeq
This shows that the variance is proportional to the square of the first derivative of the equilibrium profile. This is consistent with Schneider \cite{schneider2015physics}, where the same result was obtained with a Taylor expansion of a mean state under the assumption that the variability comes from advection of a mean state. 

\subsection{Third Moment}
 
\begin{figure}
\centering
    \includegraphics[height = 4 cm]{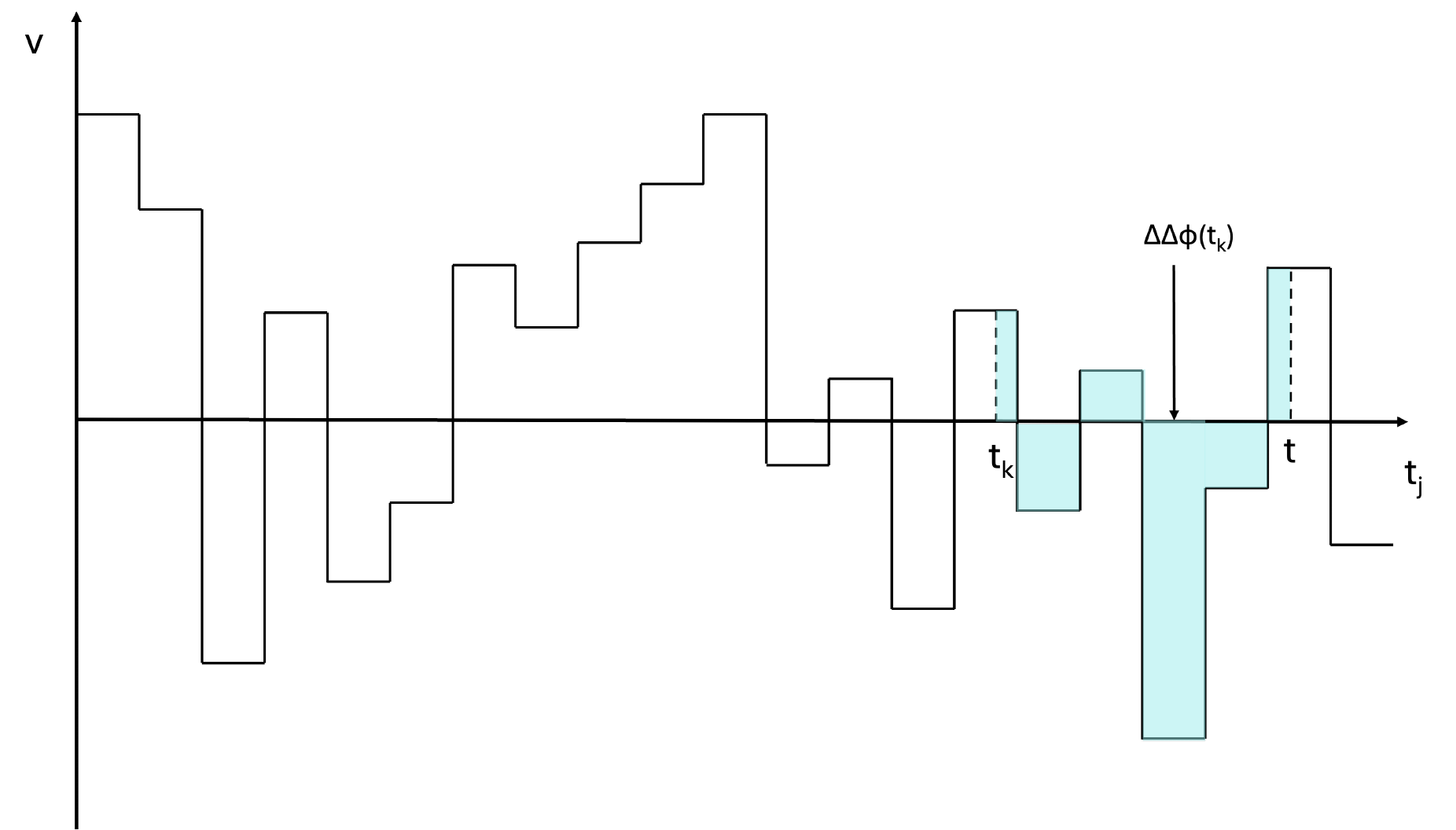}
    \caption{A graphical representation of $\xi(t_k)$. Its correlation with $v(t_k)$ is nonzero only if $t_k \le t_j \le t$.}
    \label{figure: delta_delta_phi}
\end{figure}

The third moment (skewness) $\sigma _u ^3 \mu_3$ is given by
\beq
\sigma _u^3 \mu _3 = \overline{(u-\bar u)^3},
\eeq
where
\beq
\begin{split}
&u-\bar u=- \frac 12 u_e'' \sigma _v^2 \Delta t \tau - u_e' e^{-t/\tau}\int_0 ^t v(t')e^{t'/\tau} dt' &\\
&- u_e''e^{-t/\tau}\int _0 ^t(\Delta x(t')v(t')-\Delta x (t)v(t'))e^{t'/\tau} dt' .&\\
 \end{split}
\eeq
To the lowest order in $v$, 
\beq
\begin{split}
&\sigma _u^3 \mu _3 =-3 e^{-2t/\tau} u_e^{\prime 2} \overline{M^2} \cdot \frac 12 \sigma _v ^2 \tau \Delta t u_e'' &\\
&+3e^{-2t/\tau} u_e^{\prime 2} u_e'' e^{-t/\tau} \overline{M^2P},&\\\label{eq: third_moment}
\end{split}
\eeq
where
\beq
M = \int _0 ^tv(t')e^{t'/\tau} dt' 
\eeq
and 
\beq
P =  \int _0 ^t(\Delta x(t)v(t'') - \Delta x (t'')v(t''))e^{t''/\tau} dt'' .
\eeq
Then 
\beq
\overline{M^2} = \frac 12 \sigma _v^2 \tau \Delta t e^{2t/\tau}
\eeq
and 
\beq
\begin{split}
& \overline{M^2P} = \int _{[0,t]^3} \overline{v(t_i)v(t_j)\xi(t_k) v(t_k) }e^{(t_i+t_j+t_k)/\tau} dt^3,& \label{eq: MP1}
 \end{split}
\eeq
where $\xi (t_k) = \Delta x (t) - \Delta x (t_k)$ and $\int _{[0,t]^n} dt^n$ is a shorthand notation for $\int _0 ^t \int _0 ^t \cdots \int _0^t dt^n$ (in the present case, $dt^3 = dt_idt_jdt_k$). With $\xi(t_k) = \int _0 ^{t} v(t'') dt'' - \int _0 ^{t_k} v(t'') dt'' = \int _{t_k} ^{t} v(t'') dt''$ and Eq. (\ref{eq: v}), $\overline{v(t_i)v(t_j)\xi(t_k) v(t_k) }$ 
can be written as 
\beq
\overline{v(t_i)v(t_j)[\epsilon(t_k)v(t_k) + v(t_{k+1}) + \dots + v(t)]\Delta t v(t_k) }, \label{eq: F}
\eeq
where $\epsilon(t_k)$ is defined the same way as in the discussion of the first moment. Each term in this expression is of the form $\overline{v(t_1)v(t_2)v(t_3)v(t_4)}$, and it is nonzero if and only if two pairs of correlating $v$'s are present. Furthermore, since 
\beq
\begin{split}
&E[v(t_m)^2 v(t_n)^2] - E[v(t_m)^2 ]E[v(t_n)^2] &\\
&= cov[v(t_m)^2 v(t_n)^2] = var[v(t_m)^2]\delta _{mn}&
\end{split}
\eeq
because there is no correlation between $v(t_m)$ and $v(t_n)$ when $m\ne n$, 
\beq
\begin{split}
&E[v(t_i)^2 v(t_k)^2]&\\
& =  E[v(t_i)^2 ]E[v(t_k)^2] + var[v(t_i)^2]\delta _{ik}. &\label{eq: ok?}
\end{split}
\eeq
After three integrations over $t_i, t_j$, and $t_k$ in Eq.(\ref{eq: F}), terms that come from $var[v(t_m)^2]\delta _{mn}$ are negligible compared to the terms that come from $E[v(t_i)^2 ]E[v(t_k)^2]$ due to the extra factor $(\Delta t/\tau)^2\ll 1$ compared to other terms. We also ignore the term $\epsilon(t_k)v(t_k)$, unless the correlation with this term is made with $v(t_k)$ (as opposed to $v(t_i)$ and $v(t_j)$). Hence, we find 
\beq
\begin{split}
&\overline{M^2P} = \int _{[0,t]^3}\overline{\xi (t_k) v(t_k)} \ \overline {v(t_i)v(t_j)}e^{(t_i + t_j + t_k)/\tau} dt^3 &\\
&+2\int _{[0,t]^3}\overline{\xi (t_k) v(t_j)}\  \overline {v(t_i)v(t_k)}e^{(t_i + t_j + t_k)/\tau} dt^3,&
\end{split}
\eeq
where the factor of 2 in the second term comes from the symmetry between $i$ and $j$.
Note that,
\beq
\begin{split}
&\int _0 ^t \overline{v(t_j)\xi (t_k)} e^{(t_j + t_k)/\tau} dt_j  &\\
&= \int _{t_k} ^t \sigma _v^2 \Delta t e^{(t_j + t_k)/\tau} dt_j,&
\end{split}
\eeq
as $v(t_j)$ and $\xi (t_k)$ have a nonzero correlation only if $t_k \le t_j \le t$ (see FIG. \ref{figure: delta_delta_phi}). 
Therefore,
\beq
\begin{split}
&\overline{M^2P} = \frac{\Delta t}{2} \int _0 ^t \sigma _v ^2  e^{2t_i/\tau} dt_i \int _0 ^t \sigma _v ^2  e^{t_k/\tau} dt_k \Delta t & \\
 &+ 2 \Delta t \int _0 ^t \sigma _v ^2  e^{2t_k/\tau} dt_k \int _{t_k} ^t \sigma _v ^2  e^{t_j/\tau} dt_j \Delta t &\\
 &= \frac{7}{12}\Delta t^2 \tau^2 \sigma _v^4 e^{3t/\tau}.&
 \end{split}
\eeq
To the lowest order in $\sigma_v$, Eq.(\ref{eq: third_moment}) can be written as
\beq
\begin{split}
&\sigma _u^3 \mu _3 = 3 u_e^{\prime 2} u_e '' \Delta t^2 \tau ^2 \sigma _v^2 \left(-\frac 14 + \frac {7}{12} \right) & \\
&= 4u_e ^{\prime \prime} \left(\frac 12 u_e'^{2} \Delta t \tau \sigma_v^2 \right) \cdot \left(\frac 12 \Delta t \tau \sigma_v^2 \right)&\\
&= 4 u_e '' \sigma _u^2 \sigma _x ^2 = 4 \bar u '' \sigma _u^2 \sigma _x ^2. &\label{eq: third_moment_final}
\end{split}
\eeq
The skewness is proportional to the second derivative of the equilibrium profile. Hence, in our model, the skewness is nonzero only if  $u_e$ is nonlinear in $x$. The vanishing skewness with linear $u_e$ is consistent with the idea that the variability of $u$ comes from the advection of $u_e$ (or the mean state); since the wind field is Gaussian, if $u_e$ is linear in $x$, then symmetric velocity fluctuations yield mirror-image fluctuations in $u$, resulting in a zero skewness.

\subsection{Fourth Moment}
\begin{figure}
\centering
\includegraphics[height = 8 cm]{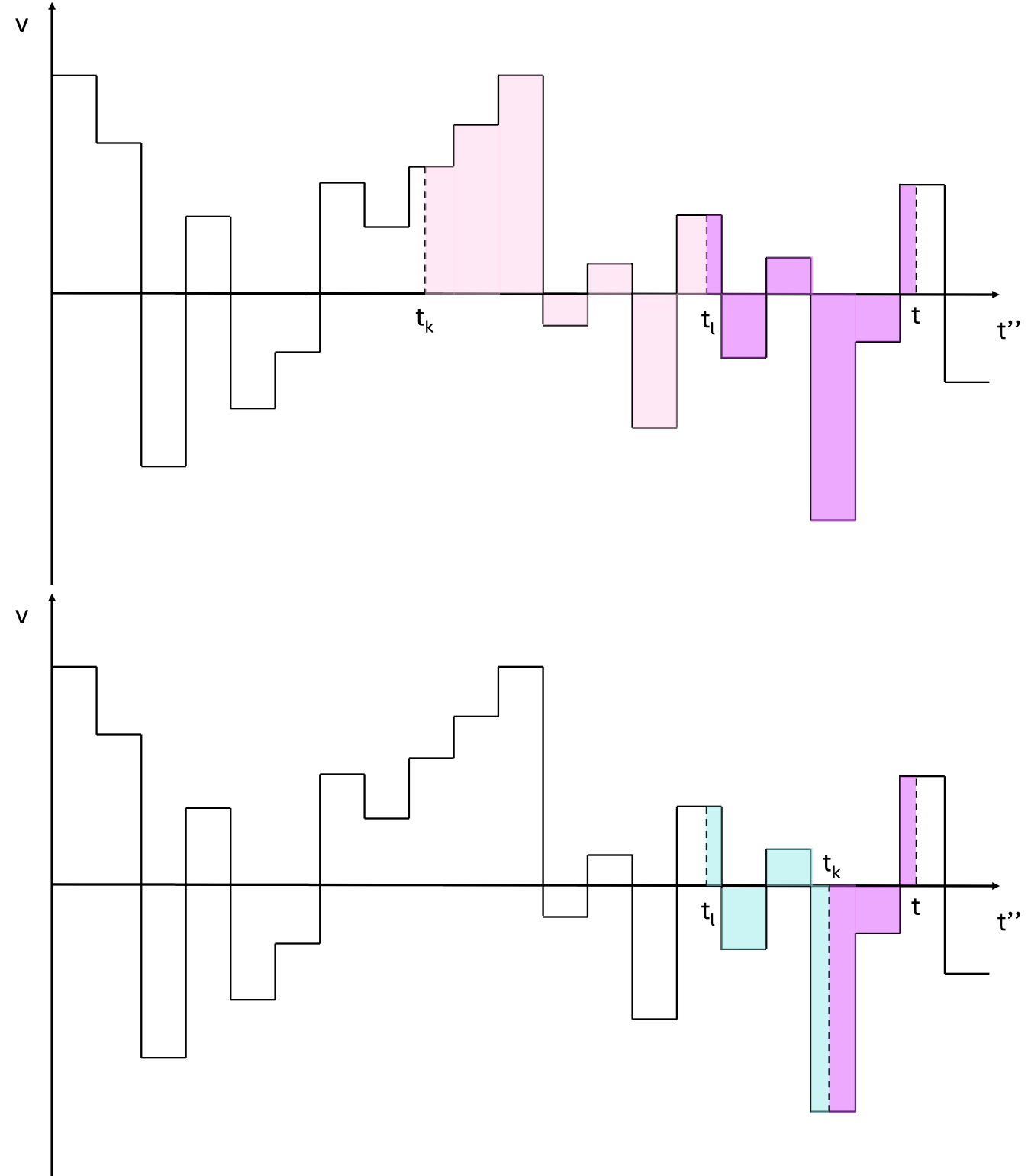}
    \caption{A graphical representation of the correlation between $\xi(t_k)$ and $\xi(t_l)$. If we fix $t_l$, their correlation is nonzero when $t'' \in [t_l, t]$ if $t_k\le t_l$, and it is nonzero when $t'' \in [t_k, t]$ if $t_k \ge t_l$. In each case, the region with a nonzero correlation is shaded with purple.}
    \label{figure: dd_phi}
\end{figure}

To calculate the fourth moment (kurtosis), it is most convenient to first express $u(t, x)$ in terms of $\bar u$ instead of $u_e$ to ensure to keep all relevant terms. Since
\beq
\begin{split}
&u_ { e } = \bar u -\bar u ^ { \prime \prime } \sigma _x ^ { 2 }, &\\ 
&u_e ^ { \prime } ( x - \Delta x ) = u _e ^ { \prime } - u_e ^ { \prime \prime } \Delta x + \frac { 1 } { 2 } u_ e ^ { (3) } \Delta x ^ { 2 },& \\ 
&u_e ^ { \prime \prime } ( x - \Delta x ) = u _e ^ { \prime \prime }  - u_ e ^ { ( 3 ) } \Delta x,&
\end{split}
\eeq
Eq. (\ref{eq: soln4}) can be written as 
\beq
\begin{split}
&u(t, x) = \bar u - \bar u'' \sigma _x ^2  &\\
&- e^{-t/\tau}\bar u ' \int v(t') e^{t'/\tau} dt' & \\
&-e^{-t/\tau}\bar u''\int \xi (t')v(t') e^{t'/\tau} dt'& \\
&+ \frac 12  e^{-t/\tau} \bar u^{(3)} \int \left( \xi(t') ^2 - 2\sigma _x ^2 \right) v(t') e^{t'/\tau} dt'.&
%
\end{split}
\eeq
Therefore, the kurtosis, evaluated to the second lowest order in $v$, is
\beq
\begin{split}
&\sigma _u^4 \mu _4 = \overline{(u-\bar u)^4} &\\
&= e^{-4t/\tau} \bar u^{\prime 4}  \overline{M^4} + 6 e^{-2t/\tau} \bar u^{\prime 2} \bar u'' \overline{M^2 N^2}&\\
&+2e^{-4t/\tau} \bar u^{\prime 3} \bar u^{(3)} \overline{M^3 L},& \label{eq: kurtosis_initial}
\end{split}
\eeq
where 
\beq
N = \sigma_x ^2 - e^{-t/\tau} \int _0^t \xi (t') v(t') e^{t'/\tau} dt' 
\eeq
and 
\beq
L = \int _0^t\left( \xi(t') ^2 - 2\sigma _x ^2 \right) v(t') e^{t'/\tau} dt'.
\eeq
To simplify Eq.(\ref{eq: kurtosis_initial}), let
\beq
D1 = e^{-4t/\tau} \bar {u}^{\prime 4} \overline{M^4},\\
D2 = 6 e^{-2t/\tau} \bar u^{\prime 2} \bar u''\overline{M^2 N^2},\\
D3 = 2e^{-4t/\tau} \bar u^{\prime 3} \bar u^{(3)}\overline{M^3 L},
\eeq
so that $\sigma _u^4 \mu _4 = D1 + D2 + D3$. Then
\beq
\begin{split}
&D1 = 3\bar u^{\prime 4} e^{-4t/\tau}\int _{[0,t]^2}\overline{v(t_i)^2} \ \overline{v(t_k)^2} \Delta t ^2 e^{2(t_i+t_k)/\tau}dt^2 & \\
&= 3\left(\frac 12 \bar u' \tau \Delta t \sigma _v ^2 \right)^2 = 3 \sigma _u ^4,& \label{eq: D1}
\end{split}
\eeq
which is clearly the zeroth-order term of kurtosis. Note that,
\beq
\begin{split}
&e^{-t/\tau}\overline{M^2N^2} = e^{t/\tau}\int _{[0,t]^2}\overline{v(t_i)v(t_j) }\sigma _x ^4 e^{(t_i + t_j)/\tau} dt^2 &\\
&-2 \int _{[0,t]^3}\overline{v(t_i)v(t_j) \xi (t_k) v(t_k)} \sigma _x ^2 e^{(t_i + t_j + t_k)/\tau} dt^3 &\\
&+ e^{-t/\tau}\int _{[0,t]^4} \overline{v(t_i)v(t_j) \xi (t_k) v(t_k) \xi (t_l) v(t_l) } e^{\sum t/\tau} dt^4,
\end{split}
\eeq
where $\sum t = t_i + t_j + t_k +t_l$. Then we find
\beq
D2 = - 22 \bar u^{\prime 2} \bar u^{\prime \prime 2} \sigma _x ^6 + 6  \bar u^{\prime 2} \bar u^{\prime \prime 2} e^{-4t/\tau} H, \label{eq: mu4_1}
\eeq

where
\beq
H =  \int _{[0,t]^4}\overline{v(t_i)v(t_j) \xi (t_k) v(t_k) \xi (t_l) v(t_l) } e^{\sum t/\tau} dt^4.
\eeq 
Just as in the previous case with four $v_m$'s, $\overline{v(t_1)v(t_2)v(t_3)v(t_4)v(t_5)v(t_6)}$ is nonzero if and only if 
there are three pairs of two equal $t_m$'s. 
Hence, considering only the dominant terms, we find
\beq
\begin{split}
H &= \int _{[0,t]^4}\overline{v(t_i)v(t_j)}\ \overline{ \xi (t_k) v(t_k) } \ \overline{\xi (t_l) v(t_l) } e^{\sum t/\tau} dt^4 & \\
&+\int _{[0,t]^4}\overline{v(t_i)v(t_j)}\ \overline{ \xi (t_k) v(t_l) } \ \overline{\xi (t_l) v(t_k) } e^{\sum t/\tau} dt^4 &\\
&+\int _{[0,t]^4}\overline{v(t_i)v(t_j)}\ \overline{ \xi (t_k) \xi (t_l)} \ \overline{ v(t_l)v(t_k) } e^{\sum t/\tau} dt^4 &\\
&+2\int _{[0,t]^4}\overline{v(t_i)\xi (t_k)}\ \overline{  v(t_j)\xi (t_l)} \ \overline{ v(t_k)v(t_l) } e^{\sum t/\tau} dt^4 &\\
&+2\int _{[0,t]^4}\overline{v(t_i)\xi (t_k)}\ \overline{  v(t_l)\xi (t_l)} \ \overline{ v(t_k)v(t_j) } e^{\sum t/\tau} dt^4 &\\
&+2\int _{[0,t]^4}\overline{v(t_i)\xi (t_k)}\ \overline{  v(t_k)\xi (t_l)} \ \overline{ v(t_l)v(t_j) } e^{\sum t/\tau} dt^4 &\\
&+2\int _{[0,t]^4}\overline{v(t_i)v(t_k)}\ \overline{ \xi (t_k) \xi (t_l)} \ \overline{ v(t_l)v(t_j) } e^{\sum t/\tau} dt^4. &\\ \label{eq: Hargon}
\end{split}
\eeq
Note,
\beq
\begin{split}
&\int \overline{ \xi (t_k) \xi (t_l) } e^{(t_k + t_l)/\tau} dt_k &\\
&= \int _0 ^{t_l} (t-t_l) \Delta t\sigma _v^2 e^{(t_k+t_l)/\tau} dt_k &\\
&+ \int _{t_l} ^t (t-t_k) \Delta t \sigma _v^2 e^{(t_k+t_l)/\tau} dt_k, &
\end{split}
\eeq
as correlation between $\xi(t_k)$ and $\xi(t_l)$ with a fixed $t_l$ is nonzero only when $t''$ in $\xi (t_k) = \Delta x(t) - \Delta x (t_k) =  \int _{t_k} ^t v(t'') dt''$ is in $[t_l, t]$ if $t_k\le t_l$, and is nonzero only when $t''$ is in $[t_k, t]$ if $t_k \ge t_l$ (see FIG. \ref{figure: dd_phi}). The factor $(t-t_k) \Delta t\sigma _v^2$ comes from the random walk considerations (i.e., variance is proportional to the number of steps times the length of each step).


Now, Eq. (\ref{eq: Hargon}) can be written as
\beq
\begin{split}
&H= (\Delta t \sigma _v^2)^3\int _0 ^t  e^{2t_i/\tau} dt_i \int _0 ^t \frac{1}{2} e^{t_k/\tau} dt_k \int _0 ^t \frac{1}{2} e^{t_l/\tau} dt_l &\\
&+(\Delta t \sigma _v^2)^3\Delta t \int _0 ^t e^{2t_i/\tau} dt_i \int _0 ^t \frac{1}{4}  e^{2t_k/\tau} dt_k &\\
&+(\Delta t \sigma _v^2)^3\int _0 ^t e^{2t_i/\tau} dt_i \int _0 ^t  (t-t_k) e^{2t_k/\tau} dt_k &\\
&+ 2 (\Delta t \sigma _v^2)^3\int _0 ^t e^{2t_k/\tau} dt_k \int _{t_l = t_k} ^t e^{t_j/\tau} dt_j \int _{t_k} ^t e^{t_i/\tau} dt_i &\\
&+ 2 (\Delta t \sigma _v^2)^3\int _0 ^t e^{2t_k/\tau} dt_k \int _{t_k} ^t e^{t_i/\tau} dt_i \int _0 ^t \frac{1}{2} e^{t_l/\tau} dt_l &\\
&+ 2 (\Delta t \sigma _v^2)^3\int _0 ^t e^{2t_l/\tau} dt_l \int _{t_k} ^t e^{t_i/\tau} dt_i \int _{t_l} ^t e^{t_k/\tau} dt_k & \\
&+ 2  (\Delta t \sigma _v^2)^3\int _0 ^{t_l} e^{t_k/\tau}(t-t_l) e^{(t_k+t_l)/\tau} dt_k \int _0 ^t e^{t_l/\tau} dt_l&\\
&+ 2 (\Delta t \sigma _v^2)^3\int _{t_l} ^t e^{t_k/\tau}(t-t_k)e^{(t_k+t_l)/\tau} dt_k \int _0 ^t e^{t_l/\tau} dt_l &\\
\end{split}
\eeq
Therefore, 
\beq
\begin{split}
& H = \Delta t ^3 \tau ^3 \sigma _v^6 e^{4t/\tau} \left(\frac 18 + \frac 18 + \frac 16 + \frac 16 + \frac1{12} + \frac 13 \right) &\\
&= 1 \cdot e^{4t/\tau} \Delta t ^3 \tau ^3 \sigma _v^6 = 8 \cdot \left(\frac 12 \tau \Delta t \sigma _v ^2 \right)^3 e^{4t/\tau} &\\
&= 8 \sigma _x ^6 e^{4t/\tau}.&
\end{split}
\eeq
Hence,
\beq
D2 = \bar u^ {\prime 2} \bar u ^{\prime \prime 2} \sigma _x ^6 (-22 + 6 \cdot 8) = 26 \bar u^ {\prime 2} \bar u^{\prime \prime 2} \sigma _x ^6.
\eeq
To simplify D3, note that 
\beq
\begin{split}
& \overline{M^3 L} = \int _{[0,t]^4}\overline{v(t_i)v(t_j)v(t_k)\xi (t_l)^2 v(t_l)}e^{\sum t/\tau} dt^4  &\\
& - 2\sigma _x ^2 \int _{[0,t]^4}\overline{v(t_i)v(t_j)v(t_k)v(t_l)}e^{\sum t/\tau} dt^4 &\\
&= Z - 2\sigma_x^2 \cdot  3 \sigma_x ^4,&
\end{split}
\eeq
where
\beq
\begin{split}
&Z = 3\int_{[0,t]^4}\overline{v(t_i)v(t_j)} \ \overline{v(t_k)v(t_l)} \ \overline{\xi(t_l)^2} e^{\sum t/\tau} dt^4& \\
&+ 3 \int_{[0,t]^4} \overline{v(t_i)v(t_j)} \ \overline{v(t_k)\xi(t_l)} \ \overline{\xi(t_l) v(t_l)} e^{\sum t/\tau} dt^4 &\\
&+ 3 \int _{[0,t]^4}\overline{v(t_i)\xi(t_l)} \ \overline{v(t_j)\xi(t_l)} \ \overline{v(t_k) v(t_l)} e^{\sum t/\tau} dt^4. & \\
 &= 3(\Delta t \sigma _v ^2 )^3 \int _0^t e^{2t_i/\tau} dt_i \int _0 ^t (t-t_l)  e^{2t_l/\tau} dt_l &\\
&+ 3(\Delta t \sigma _v ^2 )^3 \int _0^t e^{2t_i/\tau} dt_i \int _{t_l} ^t e^{t_k/\tau} dt_k \int _0 ^t \frac{1}{2} e^{t_l/\tau} dt_l &\\
&+ 3(\Delta t \sigma _v ^2 )^3 \int _0^t e^{2t_l/\tau} dt_l \int _{t_l} ^t e^{t_j/\tau} dt_j \int _{t_l} ^t e^{t_i/\tau} dt_i &\\
& = \left[\frac 38 + \frac 38 + 3\left(\frac 12 - \frac 23 + \frac 14 \right) \right] \tau ^3 \Delta t ^3 \sigma _v ^6 e^{4t/\tau} &\\
&= 1 \cdot \tau ^3 \Delta t ^3 \sigma _v ^6 e^{4t/\tau} = 8\sigma _x ^6 e^{4t/\tau}.&
\end{split}
\eeq
Therefore,
\beq
\begin{split}
&D3 = 2 \bar u ^{\prime 3}\bar u^{(3)} (8-6) \sigma _x ^6 &\\
&= 4 \bar u^{\prime 3} \bar u^{(3)}\sigma _x ^6 = 4 \bar u^{(3)} \sigma_x ^3 \sigma _u^3.&
\end{split}
\eeq
Finally,
\beq
\begin{split}
&\sigma _u^4 \mu_4 = D1 + D2 + D3 &\\
&= 3\sigma _u^4 + 26 \bar u ^{\prime \prime 2} \sigma _u^2 \sigma _x^4 + 4 \bar u ^{(3)} \sigma _u^3 \sigma _x^3, &\label{eq: fourth_moment}
\end{split}
\eeq
thus
\beq
\begin{split}
\mu_4 = 3+ \frac{13}{8} \mu_3 ^2+ 4 \bar u^{(3)} \frac{\sigma _x^3}{\sigma _u}.  \label{eq: fourth_moment_skew}
\end{split}
\eeq
Note that $\mu_3$ depends on both first and second derivatives of $\bar u$; kurtosis depends on first, second, and third derivatives of $\bar u$. If both second and third derivatives are zero, then excess kurtosis vanishes.

The calculations are more complicated with higher-order moments, but they can be performed in the same systematic manner.

\section{\label{sec: validation}Numerical Validation of the Analytical Framework}

In this section, we compare our first four moments, (\ref{eq: first_moment}), (\ref{eq: second_moment}), (\ref{eq: third_moment_final}), and (\ref{eq: fourth_moment_skew}), with a numerical calculation of the advection equation (\ref{eq: advection}). The equilibrium function $u_e$ is set to be a cosine function $u_e(x) = 2\cos((\pi/2)x)$, and the calculations are performed over the domain $0<x<1$.  The stochastic velocity $v(t)$ is set by choosing random values with a Gaussian distribution at each $t$, and Lax-Wendroff scheme is used for the integration.

While the analytical solution was tested across a broader range of parameters, we present two representative cases here: one under ideal conditions where the theory matches the numerics exceptionally well ($\Delta t/\tau = 0.05$ and $\sigma_v \Delta t /\Delta x= 0.04$) and one where the parameters are relaxed to illustrate the expected boundaries and minor deviations of the model ($\Delta t/\tau \sim 0.125$ and $\sigma_v \Delta t /\Delta x= 0.2$), both sets of values being motivated by typical conditions of the middle latitudes of the Earth's atmosphere.

\begin{figure}
    \centering
 \includegraphics[width = 4 cm]{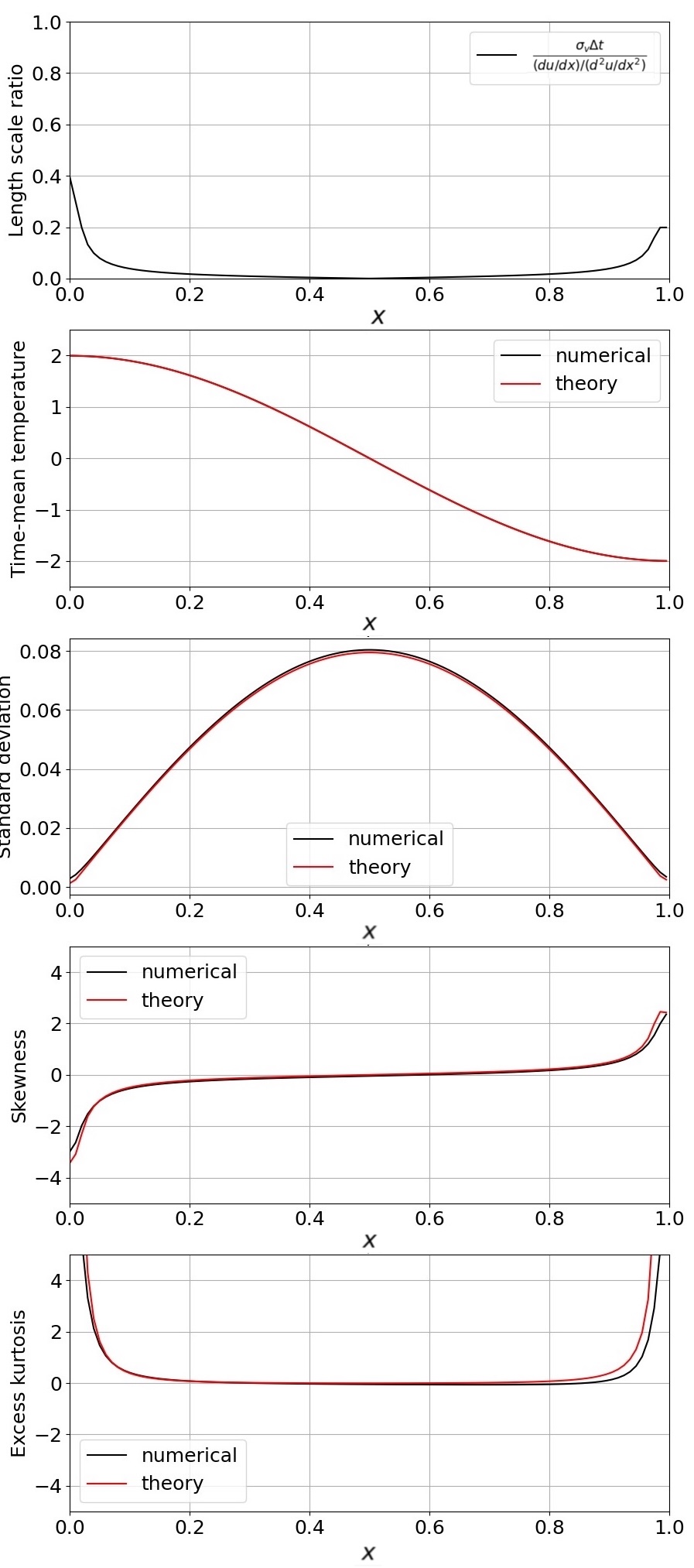}
 \includegraphics[width = 4 cm]{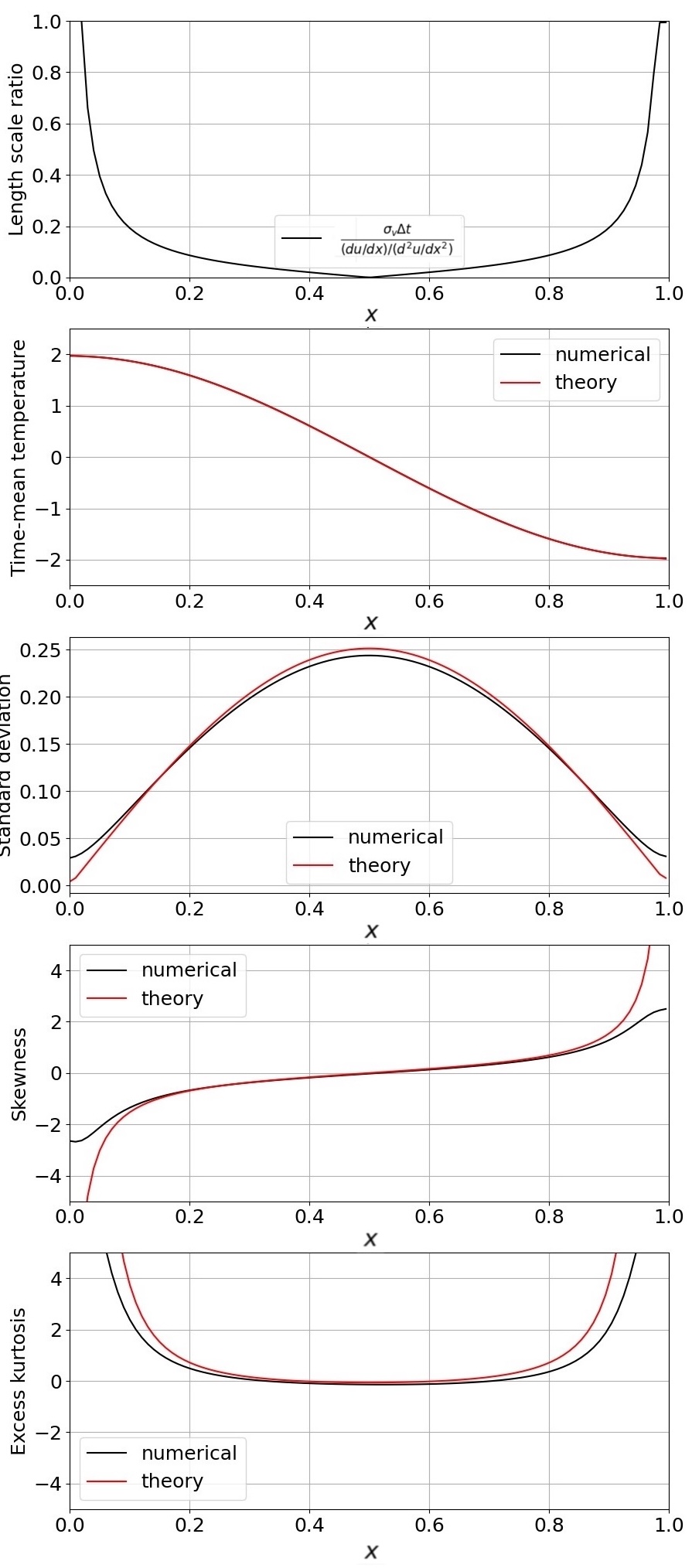}
    \caption{Comparison between our theory and numerical calculation. From top: ratio of mixing length $\sigma_v \tau_r = \sigma_v \tau$ and the ratio of $\partial u/\partial x$ and $\partial^2 u/\partial x ^2$, time-mean temperature, standard deviation, skewness, and kurtosis of $u(t,x)$. 
    Left: $\Delta t/\tau \sim 0.05$ and $\sigma_v \Delta t /\Delta x= 0.04$
     Right: $\Delta t/\tau \sim 0.125$ and $\sigma_v \Delta t /\Delta x= 0.2$. 
     As expected, analytical solutions match numerical simulations when $\Delta t/\tau \ll 1$ and $\sigma_v \Delta t/((d\bar u/dx)/(d^2\bar u/dx^2)) \ll 1$.}
    \label{figure: numerical}
\end{figure}

As our derivation in Sec.\ref{sec: PDE}, our analytical solutions of moments are for the limit where correlation time $\Delta t$ is small compared to the relaxation time $\tau$ and where the typical distance moved by the fluid within one correlation time is small compared to the length scale of the change in $d\bar u/dx$, corresponding to $\Delta t/\tau \ll 1$ and $\sigma_v \Delta t/((d\bar u/dx)/(d^2\bar u/dx^2))\ll 1$. As shown in FIG. \ref{figure: numerical}, our numerical solutions indeed match our analytical solutions for all four moments when $\sigma_v \Delta t/((d\bar u/dx)/(d^2\bar u/dx^2))$ is less than $\sim 0.2$, and the region of poor agreement is smaller with smaller $\Delta t/\tau$.

\section{\label{sec: conclusions}Conclusions}

In this work, we have presented a fully analytical derivation of the single-point statistical moments for the solution of a one-dimensional nonhomogeneous advection equation governed by a random transport velocity. The framework incorporates a stochastic Gaussian transport velocity with a finite, discrete correlation time $\Delta t$ and a nonzero forcing term with the linear relaxation approximation with its timescale being $\tau$. While our methodology is applicable to moments of any arbitrary order, we have provided explicit closed-form expressions for the first four moments in the regime where correlation time is much smaller than the relaxation timescale ($\Delta t \ll \tau$) and the characteristic distance traveled by the fluid within this correlation time is short compared to the characteristic length scale of the spatial change in the equilibrium profile. 

In contrast to the traditional approaches that rely on computationally intensive ensemble simulations, our approach yields closed-form expressions that depend exclusively on the mean profile $\bar u$, the relaxation timescale, and the statistical properties of the transport velocity. This provides a computationally efficient alternative for calculating statistical moments, where direct data-driven approaches suffer from inaccuracies with higher-order moments due to their sensitivitiy to outliers. Specifically, we show that non-Gaussian statistics arise directly from the nonvanishing derivatives of the underlying equilibrium profile: mean state $\bar u$ shifts away from the equilibrium profile $u_e$ due to the local curvature $\bar u^{\prime \prime}$, the variance scales with the square of the gradient $\bar u'$, skewness is proportional to the second derivative $\bar u ^{\prime \prime}$, and the kurtosis depends on all first three derivatives of $\bar u$. 

Beyond its computational efficiency, the primary value of our framework is that it directly connects the local temporal variability of field $u(t,x)$ at a fixed location (i.e., single-point statistical moments) to the macroscopic spatial gradients of its mean state $\bar u(x)$. Our expressions for the moments establish an ``equation-of-state" style approach that describes the average microscale behavior of $u(t,x)$ in terms of the macroscopic quantity $\bar u$.

These results are applicable to a broad spectrum of physical systems characterized by advection with random transport velocity and linear relaxation, such as localized temperature distributions or chemical concentration fluctuations. It is well-suited for multi-sector systems analysis, as our framework lets the radiative-advective components (e.g. atmosphere dynamics) be simplified, which reduces computational cost in coupled modeling frameworks. Future efforts include generalizing the theory to two or three dimensions, and incorporating spatial dependence into the stochastic velocity field to capture more complex turbulent transport regimes.

\begin{acknowledgments}
This work was supported by the U.S. Department of Energy, Office of Science, Biological and Environmental Research Program, Earth and Environmental Systems Modeling, MultiSector Dynamics under Cooperative Agreement DE-SC0022141. Any opinions, findings, and conclusions or recommendations expressed in this material are those of the authors and do not necessarily reflect the views of the funding agencies. We thank Neutrina Kircher for her unwavering support.
\end{acknowledgments}

\bibliography{citation.bib}

\end{document}